\documentclass[sn-basic]{sn-jnl}

\usepackage{graphicx}%
\usepackage{multirow}%
\usepackage{amsmath,amssymb,amsfonts}%
\usepackage{xcolor}%
\usepackage{textcomp}%
\usepackage{manyfoot}%
\usepackage{booktabs}%
\usepackage{type1cm}

\begin{document}

\title[A MOND model applied to the rotation curve of galaxies]{A MOND model applied to the rotation curve of galaxies}


\author[1,2]{\fnm{Ana} \sur{C. M. Ciqueira}}
\equalcont{These authors contributed equally to this work.}

\author*[1]{\fnm{Geanderson} \sur{A. Carvalho}}\email{gacarvalho@utfpr.edu.br}
\equalcont{These authors contributed equally to this work.}

\author[1]{\fnm{Paulo} \sur{H. Faccin}}
\equalcont{These authors contributed equally to this work.}

\author[1]{\fnm{Fabrício} \sur{Dalmolin}}
\equalcont{These authors contributed equally to this work.}

\affil*[1]{\orgdiv{Physics Department}, \orgname{Universidade Tecnológica Federal do Paraná}, \orgaddress{\street{Avenida Brasil 4232}, \city{Medianeira}, \postcode{85722-332}, \state{Paraná}, \country{Brazil}}}

\affil[2]{\orgdiv{João Manoel Mondrone State College}, \orgaddress{\street{Rua Mato Grosso 2233}, \city{Medianeira}, \postcode{85884-000}, \state{Paraná}, \country{Brazil}}}


\abstract{In this work, we propose a modified Newton dynamics (MOND) model to study the rotation curves of galaxies. The model is described by an arctangent interpolating function and it fits the rotation curves of several galaxies without invoking the presence of dark matter. We took from the literature the rotation curve data of fifteen spiral galaxies, and used it to constrain the model parameter, $a_0$, as around $5\times 10^{-10}$ m/s$^2$. This parameter is also called the acceleration constant once it gives the acceleration scale where Newton's dynamics fails. The model can be further tested in different astrophysical scenarios, such as, the missing mass problem of galaxy clusters and the accelerated expansion of the Universe, thus leading to a more robust and well constrained model.}




\maketitle

\section{Introduction}\label{sec:intro}

The enigmatic missing mass problem is a decades old, open question in modern physics \cite{Bertone2018Oct}. It started with Oort and Zwicky in the early 30s, where they have found a large discrepancy between velocities obtained from the luminosity of stars and from the virial theorem \cite{Zwicky1937Oct}. The high observed velocities could be explained if one assumes that a large amount of mass was unseen. Later in the 70s, the rotation curve of galaxies indicated also that the gravity of visible matter in the galaxies is not consistent with the high orbital speeds of the stars, so, the Newton's theory of gravitation was not enough to explain the galactic dynamics \cite{Rubin1970Feb,Rubin1976Sep,Rubin1980Jun}. Again, the problem is solved by assuming the existence of a dark halo in the galaxies. Those discoveries established the dark matter concept as the viable explanation for the missing mass problems.

Another way to explain the rotation curve of galaxies is to assume that Newtonian dynamics break down at galactic acceleration scales instead of assuming the dark matter existence \cite{Milgrom1983Jula,Milgrom1983Julb,Milgrom1983Julc}. The pioneer work of Milgrom introduced a modified Newton dynamics (MOND) which assembled important results, such as: the galaxy Keplerian, circular velocities become constant for large radial distances, and the asymptotic velocity, $v_{\infty}^4=a_0GM$, is given only by galaxy's total mass and $a_0$, which is the acceleration parameter (typically in the order of $10^{-10}$ m/s$^2$). In MOND theory, the rotation curves are well-fitted without invoking the presence of an ``invisible" matter and the Tully-Fisher relation is also naturally obtained as a consequence of the modified dynamics.

Milgrom's proposal consists of modifying the Newton's second law of dynamics by using an auxiliary function $\mu(x)$, where $x=a/a_0$, such that $a\mu=a_N$, with $a_N$ being the Newtonian acceleration. Another way of thinking is in terms of a modified gravitational field, such that $g\mu(g/a_0)=g_N$, where $g_N$ is the conventional Newtonian gravitational field, which in this case keeps the second law unchanged. The Newtonian gravitational acceleration, $g_N$, is assumed to have usual dependence on its sources and their spatial distributions. Both concepts yield to the same main conclusions. 

The function $\mu$ is constructed such that it recovers desired phenomenological aspects of rotation curves of galaxies. In that sense, MOND is not derived from first principles, which is one of the drawbacks of the theory. Accordingly, the function $\mu$ must approximate 1, for $x>>1$, and $\mu\approx x$ for $x<<1$. In the former scenario, the theory recovers the Newtonian theory in acceleration scales larger than $a_0$, in the latter case the theory provides a gravitational pull proportional to the square of $a$, and this low acceleration regime is called deep-MOND. Those constraints on $\mu$ lead to a wide range of possibilities for it, one of the must used ones is $\mu(x)=x/(1+x)$, which can be generalized to $\mu(x)=x/\sqrt[n]{1+x^n}$, for any real $n>0$.

It is worth to cite that the most accepted theory of gravity is Einstein's General Relativity (GR). GR is successful in explaining a large variety of phenomena, such as, the Mercury's perihelion advance, light deflection, frame dragging, black holes physics, gravitational redshift, gravitational lensing, binary pulsars and so on \citep{Sanders2002Sep,Will2014Dec,BuggD.V.2014Mar}. However, at small accelerations, GR recovers the Newton's theory of gravitation, so it also fails when applied to describe the dynamics of galaxies, globular clusters, and of the Universe. A challenge to Milgrom's theory is a relativistic development of it, what is called by \cite{McGaughStacy2014Apr} of a paradigm between MOND and $\Lambda$CDM, the former lacks a deeper physical derivation and the latter lacks direct detection, see also \citep{Magueijo2007Dec,Dodelson2011Dec,Kroupa2012Dec,Famaey2012Dec,Milgrom2014Jun,Iorio2015Apr,Debono2016Sep,Vishwakarma2016May,Sivaram2017Oct,BeltranJimenez2019Jul,Banik2022Jun,Duerr2023Oct}. Extensions of GR are also often used in the literature and are also an avenue to explore new insights into physical phenomena \citep{Exirifard2013Jun,Lobato2019Apr,Rocha2020May,Carvalho2020Jun,Carvalho2021Feb,Lobato2022Jun,Yousaf2023Oct,Asad2024Dec,Almutairi2024Sep,Iorio2024Dec,Bhatti2024Feb}. Recently, the existence of solitons and a model called $\kappa$-model, similar to MOND, were also successful in explaining the galactic dynamics overcoming the dark matter paradigm and supporting the Tully-Fisher relation \citep{Vukcevic2024Sep,Pascoli2024Mar}.

In this work, we focus on the study of rotation curve of galaxies by developing a new proposal for the interpolating function, $\mu$. We took a sample of 15 galaxies to test the model. The article is organized as follows: in section \ref{sec:MM} the model is highlighted, while in \ref{sec:results} the results for rotation curves are presented, finally, in section \ref{sec:dc} we give some discussion and conclusions.

\section{The arc-tangent model}\label{sec:MM}

\begin{figure}
    \centering
    \includegraphics[scale=0.7]{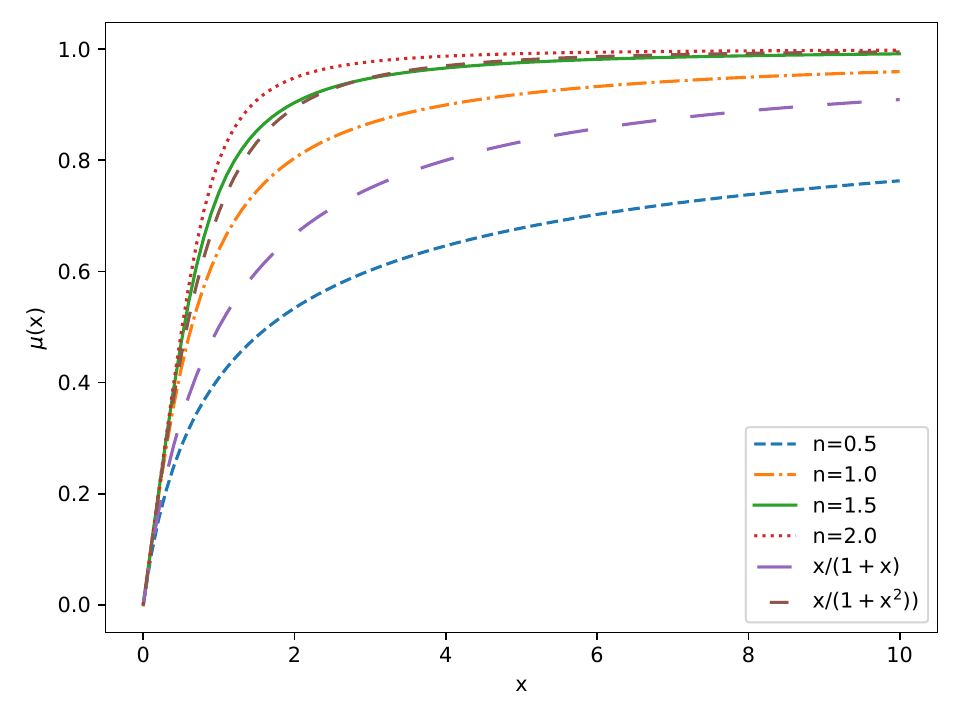}
    \caption{Interpolating function versus normalized acceleration. The arc-tangent function with $n=0.5,1.0,1.5,2.0$ is shown together with usual choices for $\mu$ within the literature. See text for more details.}
    \label{fig:interpfunc}
\end{figure}

 Considering the aforementioned constraints imposed to the interpolating function, we choose to test a new $\mu$ function, given by,
 \begin{equation}\label{arctan}
     \mu(x)= \frac{2}{\pi}\arctan\left(\frac{\pi x}{2}\right),
 \end{equation}
which can be straightforward generalized to
\begin{equation}\label{IF}
     \mu(x)= \left[\frac{2}{\pi}\arctan\left(\frac{\pi x^n}{2}\right)\right]^{1/n},
 \end{equation}
which is valid for any $n>0$. In Fig. \ref{fig:interpfunc}, we show the interpolating function Eq. \eqref{IF} for some values of $n$, and for a comparison purpose, we also show 
\begin{equation}\label{squaren}
    \mu(x)=\frac{x}{\sqrt[n]{1+x^n}}
\end{equation} 
with $n=1,2$.

From Fig. \ref{fig:interpfunc}, one can see that the arc-tangent interpolating function with $n=1.5$ resembles Eq. \eqref{squaren} with $n=2$, while the arc-tangent function with $n=1$ is between Eq. \eqref{squaren} for $n=1$ and $n=2$, so, in this case, the arc-tangent function is between the two simple functions frequently used in the literature. Here, the arc-tangent function is adopted with $n=1$. For the called ``simple" interpolating function (Eq. \eqref{squaren} with $n=1$), the values of $a_0$ that are compatible with the rotation curve of galaxies disagree with orbits of planets in the inner Solar System, while the called ``standard" interpolating function (Eq. \eqref{squaren} with $n=2$) leads to a relatively sharp transition between deep-MOND and Newtonian regimes \citep{Gentile2011Mar}. The solution is to use an improved model to intercalate between the two, such as the arc-tangent model proposed in this work.

\begin{figure}[htb!]
\hspace{-1cm}
\includegraphics[scale=0.55]{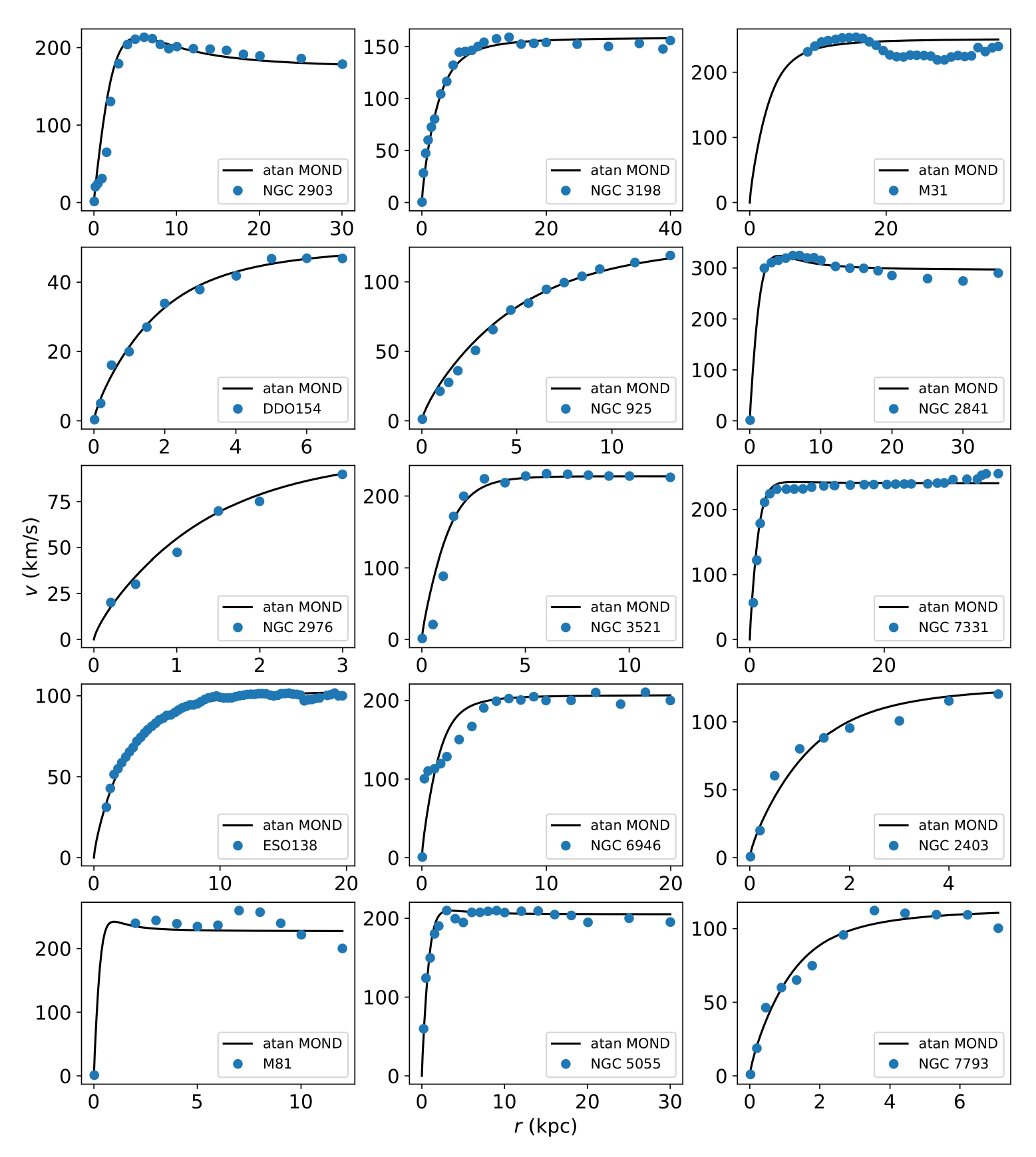}
\caption{Fits of the rotation curves of fifteen galaxies. The value of the acceleration parameter, $a_0$ is fixed as $5\times 10^{-10}$ m/s$^2$, except for NGC 2903 where the value $a_0=8\times 10^{-11}$ m/s$^2$ was adopted. The total mass and total radius of the galaxy are also used as parameters for the fits. Rotation curve data were taken from \cite{Perko2021Jul}.}\label{fig:fits}
\end{figure}

\section{Results}\label{sec:results}

We took a sample of fifteen rotation curves of galaxies from \cite{Perko2021Jul} \footnote{The fifteen galaxies belong to the NGC catalog \citep{NGCcatalog}, and between them, 14 are spiral and 1 has an irregular shape.}, and consider that the dynamics is changed according to the arc-tangent MOND. So that, the dynamics is described by 
\begin{equation}\label{g}
    \frac{GMr}{a_0(r^2+R^2)^{3/2}}=\mu x,
\end{equation}
where $M$ represents the total mass of the galaxy, $R$ is the total radius and $G$ is the Newton's gravitational constant. It is worth to cite that the model implies that the galaxy has a disk geometry with a surface matter distribution, i.e., the galaxy is supposed to have an azimuthal symmetry and a negligible thickness, this is called the Kuzmin distribution \citep{Binney2008Jan}. 

Solving Eq.\eqref{g} for each value of radial distance from the galactic center, $r$, gives a $x(r)$ curve. The solution for $x(r)$ is performed by using a Newton-Raphson method, and with $\mu$ given by Eq. \eqref{arctan}. After getting $x(r)$, the velocity as a function of $r$ is obtained by calculating
\begin{equation}
    v(r)=\sqrt{a_0xr}.
\end{equation}

The final solution for $v(r)$ must fit the observed radial velocities for a given galaxy. The total mass, $M$, and total radius, $R$, of the galaxies are used as parameters for the fits. In figure \ref{fig:fits}, we show the fitting results for the fifteen galaxies. A unique value of $a_0=5\times 10^{-10}$ m/s$^2$ was enough to fit the rotation curves of almost all the fifteen galaxies, except for NGC 2903, where the value $a_0=8\times 10^{-11}$ m/s$^2$ was adopted. The behavior of this specific galaxy could reflect the imprints of additional effects, such as a different matter distribution within the galaxy or geometric effects not included in the model. A fluctuation on the best value of $a_0$ for each galaxy is also expected because of the gravitational interaction with its vicinity \citep{Chae2020Nov}. Besides that, one of the galaxies used in our work (DDO 154) has an irregular shape and a lower surface brightness, and its rotation curve is well explained by the arc-tangent MOND. In addition, in \citep{Richtler2024Jan} authors studied an isolated elliptical galaxy and found that MOND theory is more compatible with the galactic data than dark matter models.

It is worth to note that the classical MOND theory is not derived from first principles, so, flawed as a fundamental theory, besides that, it can be considered a semi-empirical model that fits well the galactic data. Furthermore, as the arc-tangent MOND model gives reasonable results for the galaxy rotation curves it should pass additional tests, such as, the velocity dispersion in galaxy's clusters and the accelerated expansion of the universe. For example, in \citep{Kashfi2022Oct}, authors have obtained a relativistic MOND (RMOND) model to study the cosmological eras. They found that for a certain set of constraints for the theory parameters, the RMOND could reproduce the radiation-dominated, matter-dominated and de Sitter phases of the standard $\Lambda$CDM cosmology, showing the applicability of RMOND to cosmology, see also \citep{Hao2009Jun}.

\section{Discussion and conclusions}\label{sec:dc}

We have discussed in this work a new MOND interpolating function that can be considered a semi-empirical model to explain the rotation curve of galaxies. In fact, the arc-tangent function was already used in the literature as a functional to fit directly the rotational curve of galaxies as the following \cite{Miller2011Oct}
\begin{equation}\label{apjexample}
    v(r)= v_0 +v_a\frac{2}{\pi}{\rm arctan}\left(\frac{r-r_0}{r_t}\right),
\end{equation}
where $v_0$ is the central velocity, $r_0$ is the dynamic center, $v_a$ is the asymptotic velocity and $r_t$ is the turnover radius where the transition from rising to flattening of the rotation curve occurs. This describes the rotation curve itself, but not the physics behind such a behavior. However, the interpolating function must respect similar constraints, so the arc-tangent function can be applied as the basis of the MOND theory. This is the main difference between the work of \cite{Miller2011Oct} and this work. As a semi-empirical approach, the interpolating function lacks a more deep physical interpretation. There are some possibilities to explore, but those aspects of the theory are beyond the scope of this work.  

The arc-tangent functional struggles in accounting for a sharp peak found in some local, bulge-dominated rotation curves around the turnover radius \cite{Miller2011Oct}, such behavior is again observed here, particularly in the case of the NGC 2903 galaxy. The sharp peak makes a fit more difficult while keeping the acceleration constant as fixed. Introducing more parameters, such as in Eq.\eqref{apjexample}, could help to obtain a better fit while also giving more information on the properties of the galaxies. Howsoever, changing the value of $a_0$ from $5\times 10^{-10}$ to $8\times 10^{-10}$ m/s$^2$ (a 60\% increase) allowed us to well-fit the NGC 2903 data. This encompasses the applicability of the arc-tangent interpolating function as a viable model for the rotation curves of galaxies. The arc-tangent model presents an increased capability to fit the rotation curves with a single value for $a_0$. In other MOND models, the fitting procedures show some larger variability for $a_0$ \citep{Gentile2011Mar}. Also, for some galaxies, the differences in the initial slope of the rotation curves cannot be fully explained by dark matter models, however, the initial slope of rotation curves can be well-fitted in MONDian approaches.

In addition, a prediction of the so-called standard model of cosmology is that, if dark matter particles do exist, they will form massive and outspread halos around the galaxies. Hence, as a consequence, the bars of galaxies will suffer dynamical dissipation and slow down due to the dark matter halos. In \cite{Roshan2021Nov}, authors have shown that the fast rotation speeds of galactic bars is in strong disagreement with dynamical dissipation (more than 10$\sigma$ confidence level), which is a challenge to the standard model. Another finding from a similar group of authors \cite{Haslbauer2022Feb} is that there is a significant deficit between the number of intrinsically thin disk galaxies predicted in the standard model and that observed in local galaxy population from Galaxy And Mass Assembly \cite{Driver2011May,Baldry2018Mar} and Sloan Digital Sky surveys \cite{York2000Sep,Ahumada2020Jun}. Both results put the dark matter concept into serious challenge. 

Moreover, in \citep{Chae2020Nov,Chae2021Nov} authors have detected the external field effect predicted by MOND in a sample of 153 rotating galaxies from Spitzer Photometry and Accurate Rotation Curves (SPARC). \cite{Chae2023Aug} also found that the dynamics of wide binaries has better consistency with MOND than with Newton and GR theories. All these aforementioned results indicate the necessity for more research testing MOND as a viable explanation to the missing mass problems rather than research leaning to the dark matter hypothesis.

\section*{Acknowledgements}
GAC would like to thank CNPq (Conselho Nacional de Desenvolvimento Científico e Tecnológico) for financial support under process 314121/2023-4 and Fundação Araucária for financial support under NAPI ``Fenômenos extremos no Universo''. ACMC also thanks CNPq for financial support.


\bibliography{bibliography}

\end{document}